\documentclass[aps,prl,showkeys,superscriptaddress,10pt,nobibnotes,twocolumn]{revtex4-1} 

\usepackage[utf8]{inputenc}
\usepackage{dsfont}
\usepackage{amsmath} 
\usepackage{graphicx}
\usepackage{bm}
\usepackage{xcolor}

\begin{document}

\title{Magnetic monopoles and toroidal moments in LuFeO$_3$ and related compounds}


\author{Francesco Foggetti}
\affiliation{Quantum Materials Theory, Italian Institute of Technology, Via Morego 30, 16163 Genova, Italy}
\affiliation{Department of Physics, University of Genova, Via Dodecaneso, 33, 16146 Genova GE}

\author{Sang-Wook Cheong}
\affiliation{Rutgers Center for Emergent Materials and Department of Physics and Astronomy, Rutgers University, Piscataway, New Jersey 08854, USA}

\author{Sergey Artyukhin}
\affiliation{Quantum Materials Theory, Italian Institute of Technology, Via Morego 30, 16163 Genova, Italy}



\begin{abstract}
Magnetic monopoles and toroidal order are compelling features that have long been theorized but remain elusive in real materials.
Multiferroic hexagonal ferrites are an interesting realization of frustrated triangular lattice, where magnetic order is coupled to ferroelectricity and trimerization. 
Here we propose a mechanism, through which magnetic monopolar and toroidal orders emerge from the combination of  120$^\circ$ antiferromagnetism and trimerization, present in hexagonal manganites and ferrites. 
The experimentally observable signatures of magnetic monopolar and toroidal orders are identified in the inelastic neutron scattering cross section, simulated from a microscopic model of LuFeO$_3$. The non-reciprocal magnon propagation is demonstrated.
\end{abstract}


\maketitle

{\it Introduction} --
Multiferroics are a class of materials that are attracting attention due to their promise in sensing, IT and spintronic applications \cite{Khomskii2009}. The contemporary presence of properties such as ferroelectricity, electrostriction, magnetostriction, piezoelectricity, and magnetoelectricity \cite{Cheong2007} allows to build devices with interesting properties related to e.g electric transport, heat transport, information storage and manipulation. Competing magnetic interactions often lead to peculiar magnetic states and associated symmetry breaking, resulting in the interaction between magnetism and structural distortions. These states support a rich landscape of topological defects and elementary excitations that may enable novel devices \cite{Cheong2007,Ramesh2007}. Particular attention was concentrated on magnetic toroidal order \cite{Spaldin2008,Lehmann2019} and magnetic monopoles \cite{Spaldin2013,Thole2018,Meier2019} that give rise to peculiar magnetoelectricity, non-reciprocal effects \cite{Cheong18} and also lead to peculiar excitations and dynamics in spin ice \cite{Khomskii2012,Khomskii2014}.
One of the reasons for such a variety of properties in these systems is the coupling between magnetic and electric degrees of freedom, i.e. magneto-electric effect~\cite{Kimura2003}, which enables the manipulation of currents and charges through magnetic fields or engineering particular spin configurations.

\begin{figure}[ht]
\includegraphics[width=5.5cm]{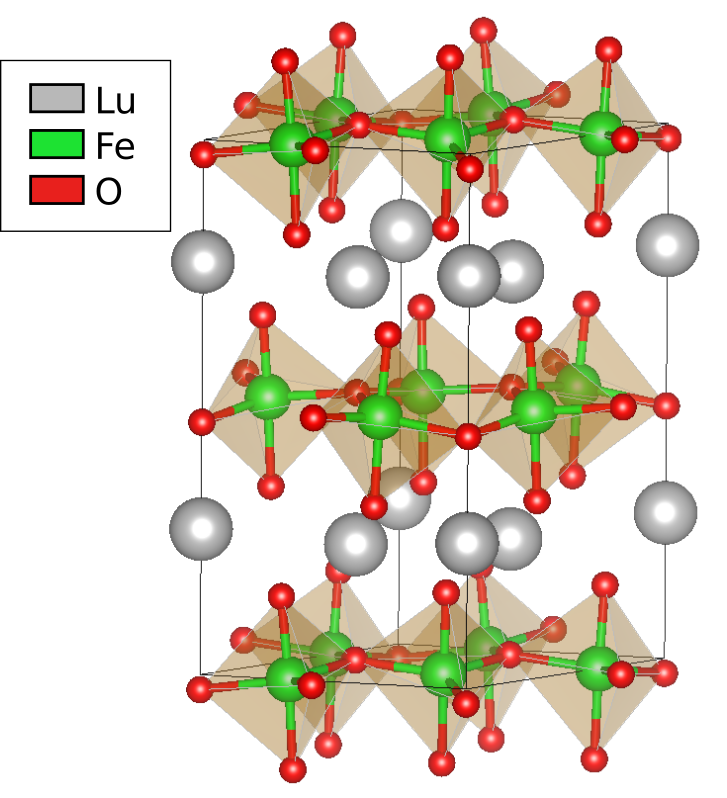}
\caption{\label{LuFeO3}The structure of LuFeO$_3$ showing triangular layers of FeO$_5$ bipyramids interspaced with Lu layers.}
\end{figure}

Antiferromagnetic triangular lattices are a compelling case of geometrically frustrated magnets because they host peculiar orders and excitations that may be utilized in the next-generation electronic devices to manipulate information without electric currents, thus reducing heat dissipation \cite{Ramesh2007}. 
Hexagonal manganites and ferrites RTMO$_3$ (TM=Mn,Fe) are multiferroics with triangular layers of magnetic ions, where unit cell-tripling buckling of bipyramids (trimerization) induces electric polarization
\cite{VanAken2004,Fennie2005}. The energy landscape has six minima along the rim of a distorted mexican hat, corresponding to alternating directions of electric polarization\cite{Artyukhin2014}. That leads to trimerization vortices, at which six trimerization domain meet, and polarization changes sign six times around a vortex core \cite{Choi2010,Artyukhin2014}.
In this work we study hexagonal LuFeO$_3$ (Fig.~\ref{LuFeO3}), a rare room-temperature multiferroic possessing weak ferromagnetic moment \cite{Wang2013,Disseler15}. Iron atoms constitute a triangular lattice and the spin configuration is determined by the competing antiferromagnetic superexchange interactions between Fe spins.

\begin{figure*}[t]
\includegraphics[width=.5\linewidth]{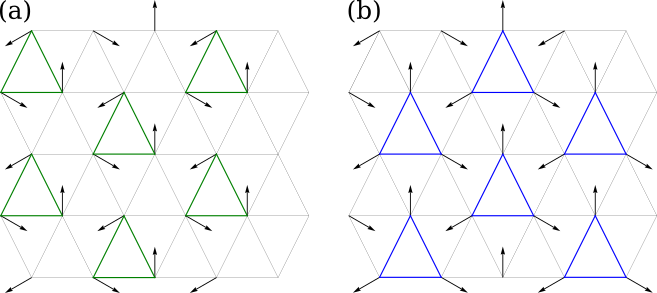}\hspace{1cm}
\includegraphics[width=.21\linewidth]{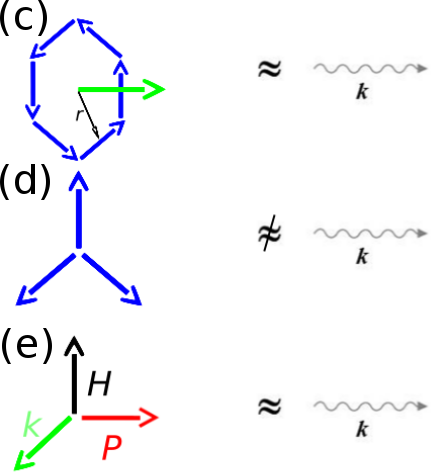}
\caption{\label{latt} 120$^\circ$ spin configuration in 2D triangular lattice. The trimers with stronger in-plane interactions are marked with colored triangles. An alternating pattern of toroidal (a) and monopolar configurations (b) emerges in this ordered phase. \label{cartoon} Symmetry considerations for non-reciprocity: c) the presence of a toroidal moment allows non-reciprocal magnon propagation; d) Monopolar moment does not support $k$-linear invariant, hence the magnon is reciprocal; e) if wavevector and the magnetic field are in the $[ab]$ plane and perpendicular to each other, and the polarization along $c$ axis is present, magnon non-reciprocity is possible.}
\end{figure*}
In order to identify the INS signatures of various states we compute magnon dispersion in LuFeO$_3$, simulate inelastic neutron scattering (INS) experiment and compare the results with existing data\cite{Leiner18}. We show signatures of magnetic monopolar and toroidal orders that are present in the lattice due to the combination of 120$^\circ$ antiferromagnetism and trimerization that lifts the cancellation of contributions from neighboring triangles. The presence of the toroidal moment gives rise to magnon non-reciprocity along the $c$ axis while the presence of monopoles is directly related to the emergence of magneto-electric (ME) effect with diagonal ME tensor \cite{Spaldin2013,Thole2016,Meier2019}. Ab-initio calculations suggest the magnetoelectric tensor in LuFeO$_3$ $\alpha_{xx}=0.26$~ps/m, $\alpha_{zz}=-3$~ps/m \cite{Ye15}.

\begin{figure*}[t]
\includegraphics[width=.7\linewidth]{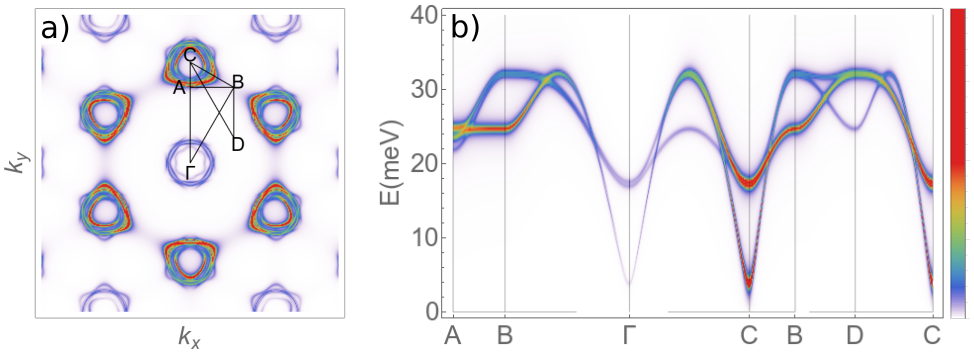} 
\caption{\label{w-slice} (a) Constant energy cut of INS cross section in $k_x,k_y$ plane at $\omega =20$~meV. The black line defines the q-space path used to plot the cross section. (b)\label{susc} Magnon spectrum along the path defined in a) for a single layer ($J'=0$) of LuFeO$_3$ with monopolar spin configuration. Colorscale encodes the INS cross-section.  
}
\end{figure*}

{\it The Model} -- 
In LuFe$O_3$ Fe has spin $5/2$ while Lu$^{3+}$ is non-magnetic. The system is composed of 2D triangular layers of Fe spins. The frustration due to antiferromagnetic exchange 
on the triangular lattice results in a 120$^\circ$ spin structure (Fig. \ref{latt}).
The single triangles that form the structure can have spins ordered in two ways, defining the toroidal and the monopolar  configurations, as shown in Fig.~\ref{latt}(a,b) respectively.
In this framework the  Hamiltonian takes the following form:
\begin{multline}
\mathcal{H}=\sum_{ij} \left(J_{ij} \vec S_i\cdot \vec S_j+\vec D_{ij}\cdot \vec S_i\times \vec S_j\right)+\\
\sum_i\left( -K(\vec S_i\cdot \vec n_i)^2+K'(S_i^z)^2-g \mu_\mathrm{B}\vec H\cdot \vec S_i\right),
\label{Ham}
\end{multline}
where the first term describes the nearest-neighbor AFM Heisenberg exchange $J_{ij}=J$ and FM interlayer exchange $J_{ij}=J'$; the second -- Dzyaloshinkii-Moriya (DM) interaction \cite{Dzyaloshinskii60,Dzyaloshinskii64,Moriya60} between the nearest neighbors in the same layer \cite{Das2014}. DM vectors were computed from experimental structural data as $\vec D_{ij}=\alpha_{DM}\vec r_{ij}\times\vec \delta$, where $\vec r_{ij}$ are the vectors connecting Fe ions, and $\vec\delta$ are the vectors connecting the middle of Fe-Fe line to the closest oxygen, with $\alpha_{DM}=0.05$~meV/\AA$^2$. The term with $K$ stands for the easy-plane anisotropy governed by the shifts $\vec n_i$ in the $ab$-plane of apical oxygens in the trimerized state. The term with $K'$ accounts for the hard-axis anisotropy perpendicular to the layers, forcing spins into the plane.
 The last term represents the interaction between the spins and an external magnetic field $\vec H$ with the gyromagnetic ratio $g=-2$. The parameter values $J=2.8$~meV, $|J'|=0.3$~meV, $H_x=2$~T, $K'=0.3$~meV, $|K|=0.68$~meV/\AA$^2$ are chosen to reproduce the experimental INS spectra \cite{Leiner18}. We started from the published parameter values \cite{Leiner18} for $J$ and $K$. In this case the spectra did not capture the experimentally observed gap at $\Gamma$ point and the energy of the plateaux between $A$ and $B$ points (Fig.~\ref{w-slice}). We then chose hard axis anisotropy and DM parameters for the simulated spectra to capture these features, and adjusted $J$ to position the plateaux at the correct energy. 
 
 Of a particular importance is the DM term that results from the displacements of oxygen ions away from the Fe-O-Fe bond center induced by trimerization and polarization modes. This polarization drives the magneto-electric effect with a diagonal or off-diagonal magneto-electric tensor for monopolar and toroidal states, respectively.

{\it Magnon non-reciprocity} --- 
In the presence of inversion or time-reversal the magnons are reciprocal, 
in the presence of trimerization, the DM interactions break inversion and allow for non-reciprocity, giving rise to peculiar transport properties. We can infer information about magnons from symmetry considerations. Table \ref{gen} shows how different observables and orders transform under the symmetry operations, Fig.~\ref{cartoon} represents the possible cases of reciprocal or non-reciprocal magnons described by our symmetry analysis. We can build the invariants entering $\omega_{k,\sigma}$ of the system by multiplying the signatures of the different quantities in the table.


\begin{table}[b]    
\begin{minipage}[t]{.62\linewidth} 
\vspace{0pt}
\begin{tabular}{|c|c|c|c|c|c|}
\hline
       &$2_{001}|(00\frac{1}{2})$&$2_{110}$&$I$&$3_z$&$T$ \\\hline
        $(\vec r_i\times \vec S_i)_z$&$+$&$-$&$-$&$+$&$-$\\
        $\vec r_i\cdot \vec S_i$&$+$&$+$&$-$&$+$&$-$\\
        $k_z$&$+$&$-$&$-$&$+$&$-$\\
        $P_z$&$+$&$-$&$-$&$+$&$+$\\
        $H_z$&$+$&$-$&$+$&$+$&$-$\\
        $(\vec k\times \vec H)_z$&$+$&$-$&$-$&$+$&$+$\\\hline
    \textrm{Phases}&&&&&\\\hline
         $A_1$&$+$&$-$&$-$&$+$&$-$\\
        $A_2$& $+$&$+$&$-$&$+$&$-$\\
         $B_1$& $-$&$+$&$+$&$+$&$-$\\
         $B_2$& $-$&$-$&$+$&$+$&$-$\\
        \hline
\end{tabular}
\hfill
\end{minipage}
\begin{minipage}[t]{0.2\linewidth}
\vspace{0pt}
\includegraphics[width=1.25cm]{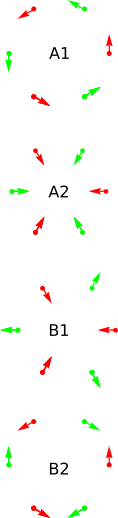}
\hfill
\end{minipage}
\caption{ \label{gen} (Left) Transformation properties under the generators of the symmetry group P6$_3$/mmc (\#194 in the International Tables). (Right) Different possible spin orders, red and blue arrows represent spins from to two different layers.}
\end{table}

\begin{figure*}[t]
\includegraphics[width=0.95\linewidth]{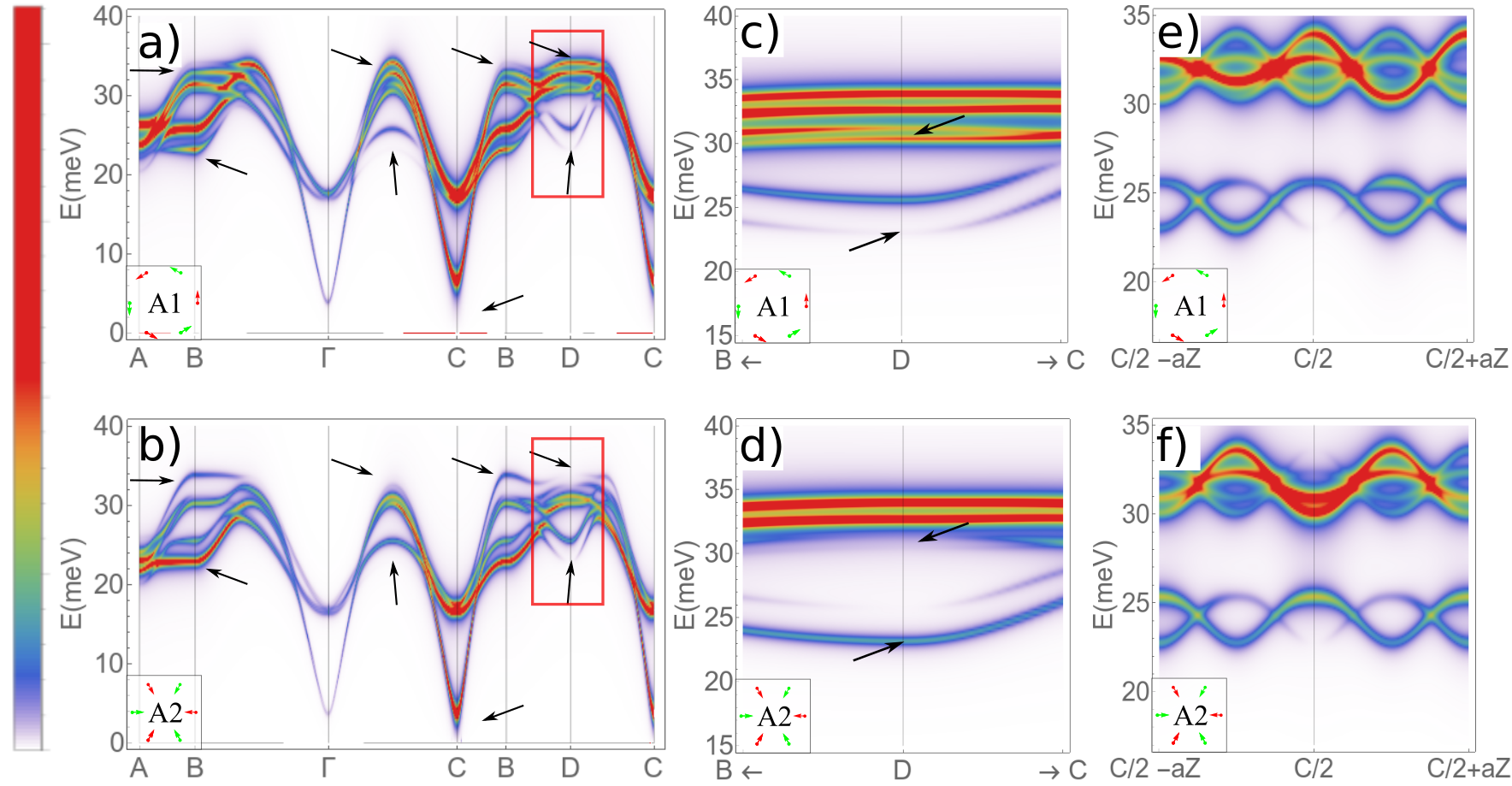}
\caption{\label{M-T} Simulated INS spectra for (a) monopolar ($K>0$) and (b) toroidal ($K<0$) states, stabilized by in-plane easy axis anisotropy. Colorscale encodes the INS cross-section. 
Differences in the intensity of the signal can be seen between the two figures. Arrows highlight the most evident ones close to $B$ and $D$ points and between $\Gamma$ and $B$;
(c,d) Close-up view of the area in red from panels (a) and (b). 
Arrows point to the differences between the two configurations.
(e,f) The INS cross-section on the peak between $C$ and $\Gamma$ points along $k_z$ in the BZ. Magnon non-reciprocity is evident in the scattering cross section for the toroidal order (e), while the magnon propagation is reciprocal in the monopolar state  (f).
}
\end{figure*}

The following situations are possible:
\begin{itemize}
    \item If the toroidal moment along $c$ axis is present then non-reciprocal spin wave propagation along $c$ axis is possible, and appears in the simulation as seen in Fig. \ref{M-T} (e). The product $k_z (\vec r_i \times \vec S_i)_z$ is in fact an invariant for the system, thus the magnon energy and the INS scattering cross-section will have contributions, proportional to $k_z$ and hence non-reciprocal: $\delta\omega\propto k_z$.

    \item The presence of monopoles alone (with $P_z=0$) does not induce non-reciprocity of spin wave propagation along $c$ axis, since $k_z(\vec r\cdot\vec S)$ is not an invariant. The absence of the linear $k$-term in $\omega$ means reciprocal propagation, as seen in the simulated spectra in Fig. \ref{M-T} (f).

    \item When $k$ and $\vec H$ are both in the $[ab]$ plane and the polarization along $c$ axis is present, non-reciprocity of the spin wave is possible. As before, the term ${k_z\cdot[\vec H \times \vec P]}$ is an invariant, meaning that a linear k-term in $\omega$ is allowed . This can generate non-reciprocity even when the system is in a monopolar configuration.
   
    \item Magneto-electric effect is possible in the presence of monopoles since the term $(\vec r\cdot\vec S)P_zH_z$ is allowed by symmetry.
\end{itemize}

We compute the magnetic susceptibility $\chi_{ij}(\omega,k)$ using linear spin wave theory as described in the Supplementary, and use it to evaluate the INS intensity. For a non-polarized neutron beam the INS cross section due to dipole-dipole interactions with neutrons is given by  \cite{Jensen1991}
\begin{equation}
\frac{d^2\sigma}{dEd\Omega}\sim\sum_{ij}\left(\delta_{ij}-\frac{k_ik_j}{k^2}\right)\chi_{ij}    
\end{equation}

The imaginary part of susceptibility gives the magnon spectral function and represents all possible magnon excitations in the material. The resulting simulated INS cross section along the path, connecting high-symmetry $k$-points marked in Fig.~\ref{w-slice}, is shown in Fig.~\ref{susc} for a single layer of LuFeO$_3$.

Repeating these calculations for toroidal and monopolar orders we obtain the spectra shown in Fig.~\ref{M-T}, and identify differences in the INS plots between monopolar and toroidal configuration. The differences are subtle in the in-plane dispersion plotted in Fig.~\ref{M-T}(a-b). However, they are evident in the $k_z$ dispersion, Fig.~\ref{M-T}(e-f), as predicted by the symmetry analysis. These differences can therefore be used to identify monopolar and toroidal orders in the future experiments. Supplementary Fig.~\ref{S1},~\ref{S2} shows how these features depend on the strength of DM interactions in all phases.

The toroidal and monopolar orders are stabilized by manipulating the easy axis anisotropy term of Hamiltonian (\ref{Ham}) and the interlayer coupling $J'$. The reversal of sign of $K$ in the model turns the easy direction into a hard one, thus rotating the easy axis by 90$^\circ$ in the $[ab]$ plane, while reversing the $J'$ term allows for ferromagnetic or antiferromagnetic  interlayer coupling  hence allowing to select the phase of interest.
Panels (c-d) of Fig.~\ref{M-T} 
show a close-up view near $D$ point, where the differences between the two cases are evident.

{\it Quadrupole contributions} --- 
In addition to magnetic monopoles, dipoles and toroidal moments, an extra term due to quadropolar moment appears in the multipole expansion of the vector potential $\vec{\mathcal{A}}$ at the same order as the toroidal moment \cite{Spaldin2008}.  
While the monopole is the zero order term, the toroidal moment and the quadrupolar moment both enter the expansion in the second order  
\begin{equation}
    \begin{split}
        &\langle {\mathcal{A}}^{(2)}_{quad}\rangle_i=-\epsilon_{ijk}q_{kl}\partial_j\partial_l\frac{1}{R}\\
    &\langle \mathcal{\vec A}^{(2)}_{tor}\rangle=\nabla(\vec t\cdot\nabla)\frac{1}{R}+4\pi\,\vec t\,\delta(\vec R)
    \end{split}
\end{equation}
where $q$ and $\vec t$ are the quadrupolar and toroidal moments, 
with $\vec t=-\frac{1}{2}g\mu_B\sum_\alpha  \vec r_\alpha \times \vec{S}_\alpha $, and 
\begin{equation}
q_{ij}=-\frac{g\mu_B}{2}\sum_\alpha({S}_{\alpha i}r_{\alpha j}+{S}_{\alpha j}r_{\alpha i}).
\label{quad}
\end{equation}
Here $g$ is the gyromagnetic factor and $\mu_B$ -- the Bohr magneton.
The three-fold symmetry only allows for $q_{zz}$ to be non-zero.
As the symmetry analysis shows, the toroidal moment contributes to magnon non-reciprocity. It's interesting to study if the quadrupolar moment contributes to the non reciprocity too, {\it i.e.} if there exists an invariant, linear in $\vec k$, that contains $q$. In the absence of external fields the tensor $q_{ij}$ can only be contracted with the vectors $\vec k$ and $\vec P$, and since $\vec P$ is along (001) the only possible combination is $k_iq_{i3}P_3$. Using Eq. (\ref{quad}), we verify that the components $q_{i3}$ are present in the $B_1$ phase but are small compared to the contributions of the toroidal moment in $A_1$ phase.
{\it Conclusions --- }
We presented the mechanism, through which magnetic monopolar and toroidal orders emerge from the combination of 120$^\circ$ antiferromagnetism and trimerization, present in hexagonal manganites and ferrites. Symmetry considerations regarding the non-reciprocal propagation of magnons are presented and corroborated by the simulations, based on a realistic microscopic model and the spin wave approximation. The simulated INS spectra allow to discriminate between monopolar and toroidal orders. We hope the results could pave the way to manipulating magnetic monopoles in hexagonal manganites and ferrites. The effects could be useful in magnon-based devices and magnonic circuits, utilizing monodirectional magnon propagation.

SWC was supported by the DOE under Grant No. DOE: DE-FG02-07ER46382.


\bibliography{spinlatt}



\widetext
\newpage
\renewcommand{\thefigure}{S\arabic{figure}}
\setcounter{figure}{0}
\begin{center}
\textbf{\large SUPPLEMENTARY INFORMATION}
\end{center}
\section{\label{SWtheory}Calculation of the inelastic neutron scattering cross section}
In order to compute INS intensity, Hamiltonian (\ref{Ham}) is expanded to the second order in the deviations of spherical angles of classical spins $
\vec S_{i}(\theta_i,\phi_i)$ 
 from their ground state values,
$
\theta_{i}=\theta_{i0}+\alpha_{i},\:\phi_{i}=\phi^n_i+\beta_{i}
$.
$\phi^n_i$ is governed by the easy axis direction $\hat n_i$. The dynamics of $(\alpha_i,\beta_i)$ near the energy minimum are governed by Hamilton equations,
\begin{equation}
\sin\theta_{i0}\,\dot{\alpha_{i}}=-\frac{\partial \mathcal{H}}{\partial \beta_{i}}\qquad
\sin\theta_{i0}\,\dot{\beta_{i}}=\frac{\partial \mathcal{H}}{\partial \alpha_{i}}
\label{EQM}
\end{equation}
In the Fourier space the Eq. (\ref{EQM}) take the form of an eigenvalue problem,
\begin{eqnarray}
&&\left(A-i\omega\mathds{1} \right)\begin{pmatrix} \alpha_k\\\beta_k\end{pmatrix}=0,\:A =\left(\begin{array}{cc}
 -\partial_{\beta_i\alpha_j}& -\partial_{\beta_i\beta_j}\\
\partial_{\alpha_i\alpha_j}& \partial_{\alpha_i\beta_i}
 \end{array}\right)\mathcal{H},\\
&&\beta_{j}=e^{ik r_j-i\omega t}\beta_{k},\qquad \alpha_{j}=e^{ik r_j-i\omega t}\alpha_{k}
\end{eqnarray}
 where $A$ is the $2n\times 2n$ matrix of the second derivatives of the Hamiltonian with respect to $\alpha_k$, $\beta_k$ and $n$ is the number of spins in the unit cell.
The neutron beam used in INS is modelled with an external time-dependent magnetic field $\vec{h}$. 
The equations of motion describing the steady-state dynamics driven by the neutron beam now take the form:
\begin{equation}
\left(A-i\omega\mathds{1} \right)\begin{pmatrix}
\alpha_k\\\beta_k\end{pmatrix}{e^{-i\omega t}}=\begin{pmatrix}
h_\alpha\\ h_\beta\end{pmatrix}{e^{-i\omega t}},
\label{extH}
\end{equation}
with $h_\alpha$ and $h_\beta$ being the terms of $\vec S\cdot \vec h$ linear in $\alpha_i$ and $\beta_i$. 
The steady-state response appears at the frequency $\omega$ of the oscillating magnetic field associated with neutrons INS experiments. 
After solving Eq.~\ref{extH} for all $\omega$ we express $\alpha_k$ and $\beta_k$ as $\omega$ dependent and find the magnetic susceptibility tensor as
\begin{equation}
\chi_{ij}(\omega,k)=\frac{g\mu_B}{V}\frac{\partial {S}^{(i)}_{k}}{\partial h^{(j)}}(\alpha_k,\beta_{k\pm Q}),
\end{equation}
where $Q$ is the wave vector of the spin texture.
For a non-polarized neutron beam the INS cross section due to dipole-dipole interactions with neutrons is given by  \cite{Jensen1991}
\begin{equation}
\frac{d^2\sigma}{dEd\Omega}\sim\sum_{ij}\left(\delta_{ij}-\frac{k_ik_j}{k^2}\right)\chi_{ij}    
\end{equation}
\section{Inelastic neutron scattering spectra for different phases and their dependence on the Dzyaloshinskii-Moriya interaction strength}
\begin{figure}[h]
\includegraphics[width=.9\linewidth]{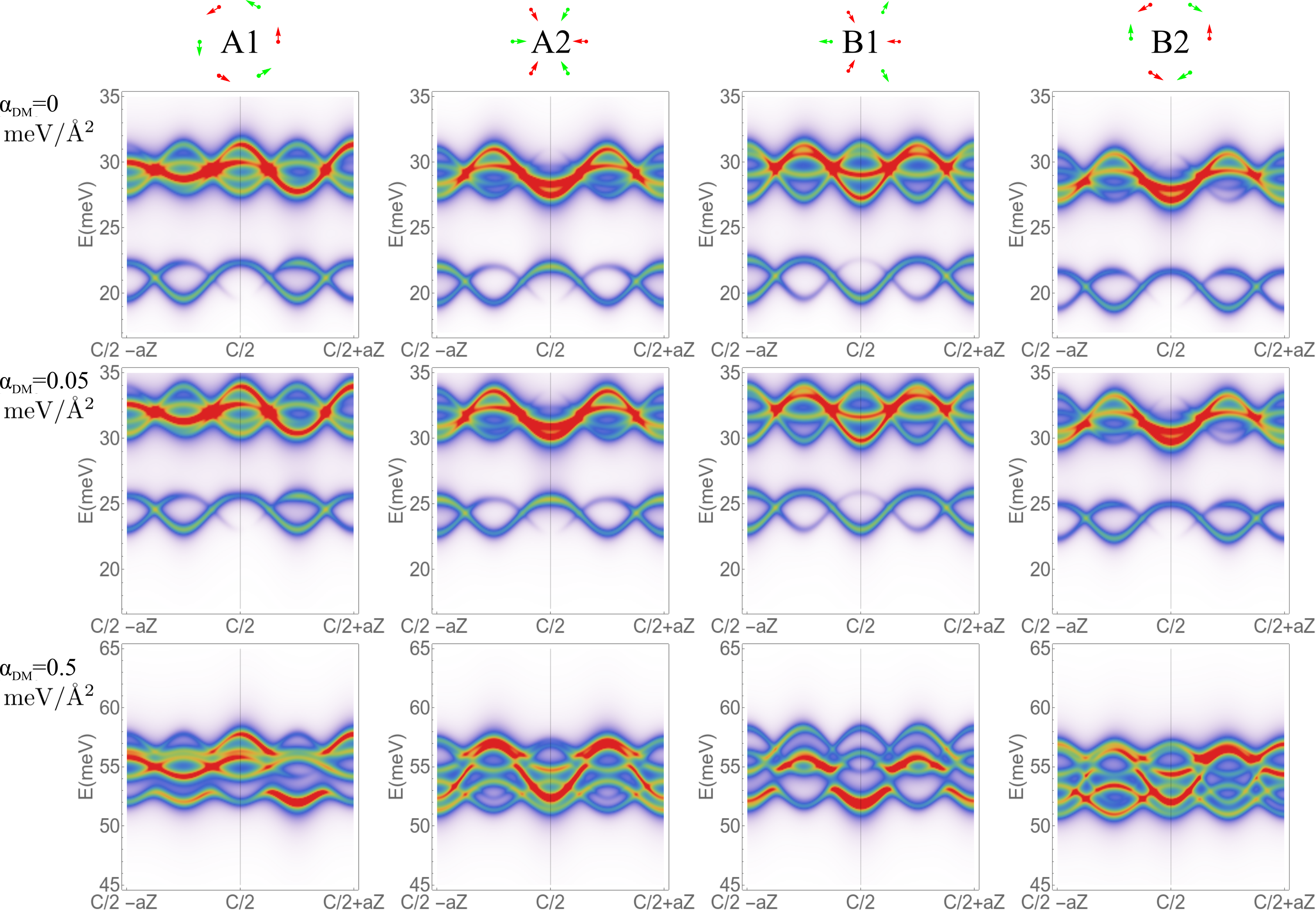}
\caption{\label{S1}INS cross-section on the peak between $C$ and $\Gamma$ points along $k_z$ in the BZ for different phases and for different values of Dzyaloshinskii-Moriya interaction. The values of the model parameters are $J=2.8$~meV, $|J'|=0.3$~meV, $h_x=2$~T, $K'=0.3$~meV, $|K|=0.68$~meV/\AA$^2$.}
\end{figure}

\begin{figure}[h]
\includegraphics[width=\linewidth]{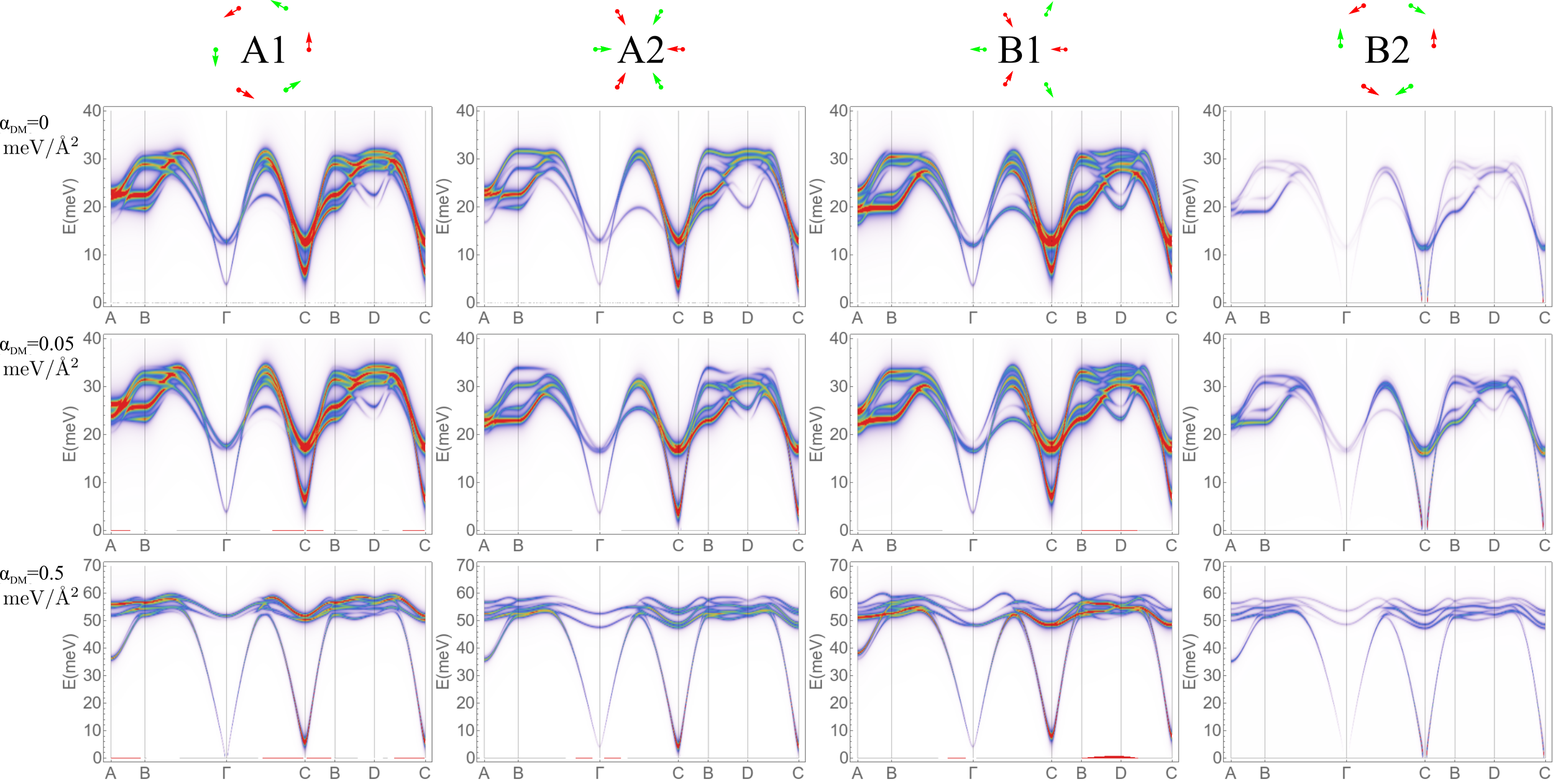}
\caption{\label{S2}Simulated INS spectra for different phases and different values of Dzyaloshinskii-Moriya interaction. 
The values of the model parameters are $J=2.8$~meV, $|J'|=0.3$~meV, $h_x=2$~T, $K'=0.3$~meV, $|K|=0.68$~meV/\AA$^2$.}
\end{figure}

Fig.~\ref{S1},~\ref{S2} illustrate the effect of Dzyaloshinskii-Moriya interactions on the INS spectra in different magnetic phases. They show how DM increases the spin wave bandwidth and enhances the non-reciprocal INS signatures in $A_1$ and $B_2$ phases, although the non-reciprocity is already evident at zero DM strength in $A_1$ phase. With increasing DM, the spins deviate slightly from the 120$^\circ$ configuration. The two sets of bands, separated at zero DM strength, merge towards $\alpha_{DM}=0.5$~meV/\AA$^2$, as seen in Fig.~\ref{S1}. The bandwidth of the overall dispersion in the hexagonal plane increases with $\alpha_{DM}$, as seen in Fig.~\ref{S2}. The dispersion along $c$ axis differs significantly for all phases, and therefore be used in order to distinguish them.

\end{document}